\documentclass{elsart}
\journal{Nuclear Physics B}

\usepackage{epsfig}
\usepackage{rotate}

\newcommand{\numucc}{$\nu_{\mu}~CC~$}
\newcommand{\anumucc}{$\bar\nu_{\mu}~CC~$}
\newcommand{\numunc}{$\nu_{\mu}~NC~$}

\newcommand{\bp}{$Bp~$}
\newcommand{\bpim}{$B\pi^{-}$~}

\newcommand{\bmath}{\begin{displaymath}}
\newcommand{\emath}{\end{displaymath}}
\newcommand{\be}{\begin{equation}}
\newcommand{\ee}{\end{equation}}

\begin{document}
\begin{frontmatter}

\title{\boldmath {A study of backward going $p$ and $\pi^{-}$ in 
\numucc interactions with the NOMAD detector }}
\collab{NOMAD Collaboration}
\author[Paris]             {P.~Astier}
\author[CERN]              {D.~Autiero}
\author[Saclay]            {A.~Baldisseri}
\author[Padova]            {M.~Baldo-Ceolin}
\author[Paris]             {M.~Banner}
\author[LAPP]              {G.~Bassompierre}
\author[Saclay]            {N.~Besson}
\author[CERN,Lausanne]     {I.~Bird}
\author[Johns Hopkins]     {B.~Blumenfeld}
\author[Padova]            {F.~Bobisut}
% ...

\address[LAPP]           {LAPP, Annecy, France}                               
\address[Johns Hopkins]  {Johns Hopkins Univ., Baltimore, MD, USA}            
% ...
\address[CERN]           {CERN, Geneva, Switzerland}                          
\address[Lausanne]       {University of Lausanne, Lausanne, Switzerland}      
% ...
\address[Padova]         {Univ. of Padova and INFN, Padova, Italy}            
\address[Paris]          {LPNHE, Univ. of Paris VI and VII, Paris, France}    
% ...
\address[Saclay]         {DAPNIA, CEA Saclay, France}                         
% ...

\clearpage

\begin{abstract}
\noindent Backward proton and $\pi^-$ production has
 been studied in \numucc interactions with carbon nuclei. 
 Detailed analyses of the momentum distributions, of the production rates,
 and of the general features of events with a backward going particle,
 have been carried out in order to understand the mechanism producing 
 these particles. The backward proton data have been compared with the 
 predictions of the reinteraction and the short range correlation models.
\end{abstract}

\begin{keyword}
Neutrino interactions, cumulative production, intranuclear cascade,
short range correlations
\PACS 13.15.+g \sep 13.85.Ni
\end{keyword}
\end{frontmatter}

\section{Introduction}
\label{intro}
It is a well established experimental fact that in the high energy 
interactions off nuclei there are particles emitted backwards, with 
respect to the beam direction, which have energies not allowed by the 
kinematics of collisions on a free and stationary nucleon.
Backward going protons are commonly observed while, in absence of
nuclear effects, their production is forbidden. High energy mesons,
whose production in the backward direction is only allowed up to a given 
momentum, are detected also, at momenta above such a limit.
Since a long time \cite{baldin} this effect has been used 
as a powerful tool to investigate nuclear structure. The models 
proposed to explain the origin of these particles (also called
cumulative in the literature) can be divided essentially into two
categories: models based on the intranuclear cascade mechanism
and models based on the cumulative effect of groups of correlated 
nucleons/quarks.\\
\noindent In the {\em intranuclear cascade models}, the
 production of particles in the kinematically forbidden region
(KFR) can be seen as the result of
multiple scattering and of interactions of secondary hadrons,
produced in the primary $\nu$-nucleon collision, with the other nucleons 
while they propagate through the nucleus \cite{kopel}. 
The reinteraction or intranuclear cascade (INC) models usually rely on 
Monte Carlo methods to make their predictions \cite{ranft}, \cite{ferrari-ranft}. 
The importance of the intranuclear cascade mechanism is that it can provide 
information on the space--time evolution of the hadronization process. 
Experimentally one observes that the cascade is restricted to slow particles 
only, while the fast ones do not reinteract inside the nucleus.
The currently accepted explanation for this effect is the ``formation zone''
concept \cite{stodolsky}-\cite{eliseev}. This is the 
distance (or the time) from the production point which is required for 
the secondary hadrons to be ``formed'', i.e. to be able to interact
as physical hadronic states. Since the distance/time required, 
due to the Lorentz time--dilation factor, 
is proportional to the energy of the secondary, 
the INC process is restricted to slow hadrons which have formation 
lengths smaller than the nuclear radius.\\
An advantage of neutrino (or more generally lepton)
induced interactions with respect to hadronic processes is the fact that
the projectile interacts only once, avoiding the complications
related to the projectile rescattering in the target. Natural drawbacks
are the facts that the equivalent of the projectile energy
in  the hadronic processes, i.e. the hadronic jet energy  in the leptonic 
processes, is not fixed but varies from event to event and suffers from 
systematic uncertainties, especially related to the reconstruction of neutrals.\\
Another feature of the neutrino--nucleus interaction is the fact that, due to 
the extremely small neutrino--nucleon cross-section, the interaction can 
practically take place anywhere in the nucleus. This is not the case in
hadron--nucleus (and also photon--nucleus) interactions where, due to the large
cross--section, the interaction takes place essentially on the first 
nucleons along the particle path, i.e. on the surface of the nucleus. 
As a consequence not all the nucleons participate in the scattering and 
the total hadron--nucleus cross--section is smaller than $A$ (the number 
of nucleons in the nucleus) times the total hadron--nucleon cross--section. 
This effect is known as {\it shadowing} \cite{arneodo}.\\
In the {\em correlated nucleon/quark models}
the backward particles are produced in the collisions off 
structures with mass larger than the mass of the nucleon. 
These structures are formed,
at small interparticle distance, under the action of the short range
part of the nuclear force. They may either be described as fluctuations of the
nuclear density \cite{burov} or as clusters of a few correlated nucleons 
\cite{fs1}-\cite{fujita2}. In any case 
these structures represent the effect of gathering two or more nucleons in
small volumes with a radius of the order of $0.5 \div 0.7~fm$. 
The nucleons in these structures can acquire high momenta and 
the fast backward going particles can be seen as a direct manifestation 
of the  high momentum tail of the Fermi distribution. 
For instance, the classical mechanism of Fermi motion of nucleons
inside the nucleus can explain backward proton production only up
to about $300~MeV/c$, well below the observed limit.
Nucleons can also loose their identity in these structures and larger masses
can be reached if quarks in neighbouring nucleons stick
together, forming multiquark clusters \cite{carlson,multiq}.
The momentum distribution of quarks in the cluster is, 
in this case, responsible for the observed spectra of cumulative particles.\\
\noindent These two classes of models  are not mutually exclusive. 
Indeed some features of the data can be explained by both, and 
the experimentally observed production in  the backward hemisphere
\cite{bosveld} can have contributions from both mechanisms.\\
\noindent While the production of particles in the KFR has been investigated
over a very wide range of incident energies and on many different nuclear targets
using hadron beams \cite{fredriksson}, the data on backward particle
 production in $\nu~(\bar\nu)$ induced reactions are far less abundant.
Up to now only five bubble chamber experiments have studied backward protons 
(and in only one experiment backward pions as well) using bubble chambers filled 
with deuterium \cite{matsinos}, heavy liquid fillings \cite{berge}-\cite{matsinos}
and in one case a hybrid emulsion--bubble chamber technique \cite{dayon}.\\
Early hadronic experiments had found that the invariant cross-section
for backward particle production can be parametrized as $e^{-BP^2}$ where $P$
is the particle momentum. It was also found that the slope parameter $B$
is almost independent of the type of incident particle, its energy and target
type, a fact known as ``nuclear scaling''. Neutrino experiments have confirmed 
these properties.\\
Here we present a detailed study of backward going protons (\bp) and
backward going $\pi^{-}$ (\bpim) produced in charged current neutrino 
interactions in the NOMAD detector. In Sec.~\ref{nomad} 
and Sec.~\ref{data} we describe the experimental apparatus and the data treatment.
In Sec.~\ref{invslope} we report the invariant cross section
distributions of \bp and \bpim. In Sec.~\ref{kine} we discuss the kinematical
properties of events with a backward particle. In Sec.~\ref{bpy} we give the
production rates and study their dependence on the atomic number using the 
results of other neutrino and hadron experiments as well. 
Finally in Sec.~\ref{data&model} we compare the data with the predictions 
of theoretical models.

\section{NOMAD detector and neutrino beam}
\label{nomad}
\begin{figure}[htb]
\begin{center}
\vskip -0.5cm
   \epsfig{file=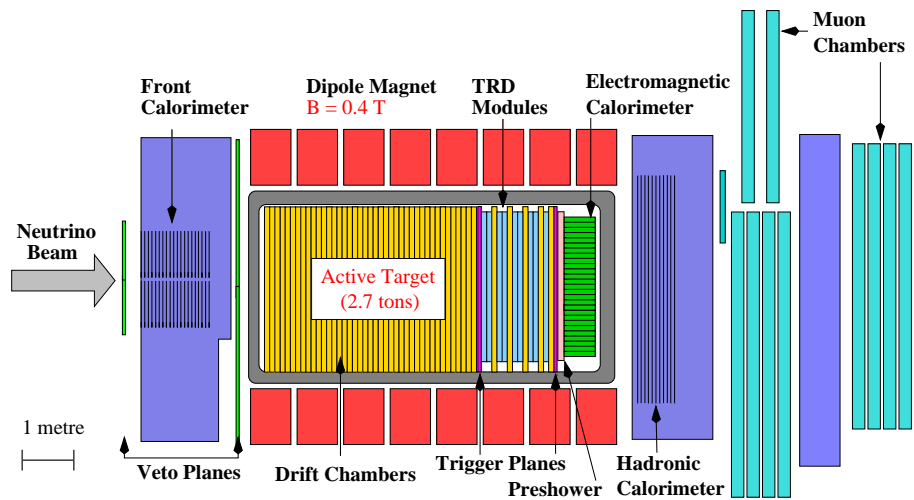,angle=-90,width=120mm}
\end{center}
\caption{{\it Side view of the NOMAD detector. }}
\label{detector}
\end{figure}

For the study of backward particles presented here, the most important 
component of the NOMAD detector is the active target. 
Its main features are highlighted here, 
while a detailed description of the full detector (shown in Fig.~\ref{detector}) 
can be found in Ref. \cite{nom1} and \cite{nom2}. \\
The target was designed with two conflicting requirements. It had to be 
as light as possible in order to minimize photon conversions, 
multiple scattering and secondary hadronic interactions, 
and it had to be as heavy as possible in order 
to produce a large number of neutrino interactions.\\
This conflict was resolved using an active target (2.7 tons) of 44
drift chambers (DC), $3\times3 m^{2}$ each \cite{nomad_dc}, 
perpendicular to the beam axis. The target mass is provided by the chamber 
structure having an average density of $0.1~g/cm^{3}$.
The drift chamber composition is reported in Table~\ref{table:dc_comp}, 
where it can be seen that carbon and elements with nearby atomic numbers
represent over 90\% of the total weight.
For  this reason we consider this study as a measurement of backward production
in $\nu_{\mu} C$ interactions.  
The chambers are placed inside a $0.4~T$ magnetic field and provide a momentum 
resolution which can be parametrized as:

\begin{displaymath}
{\sigma_{P} \over P} = {0.05 \over \sqrt L} \oplus {0.008 P \over \sqrt{L^5} }
\end{displaymath}

\noindent where $L$ is the track length in meters and $P$ the momentum in $GeV/c$.
The first term is the contribution from multiple scattering and the second 
comes from the single hit resolution.

\begin{table}[htb]
\centering
\begin{tabular}{crrrr}
\hline
Element & Z   & Weight (\%)  & p (\%) & n (\%) \\ 
\hline
H/D & 1  & 5.14  & 5.09  & 0.05 \\
C   & 6  & 64.30 & 32.12 & 32.18 \\
N   & 7  & 5.92  & 2.96  & 2.96 \\
O   & 8  & 22.13 & 11.07 & 11.07 \\
Al  & 13 & 1.71  & 0.82  & 0.89 \\
Si  & 14 & 0.27  & 0.13  & 0.14 \\
Cl  & 17 & 0.30  & 0.14  & 0.16 \\
Ar  & 18 & 0.19  & 0.09  & 0.10 \\
Cu  & 29 & 0.03  & 0.01  & 0.02 \\ \hline
&   &       & 52.43 & 47.56 \\ \hline
\end{tabular}
\caption{{\it NOMAD drift chamber composition, showing the
 proportions by weight of atoms, and of protons and neutrons. As can be 
seen the target consists mainly of carbon and is practically isoscalar.}}
\label{table:dc_comp}
\end{table}

\noindent The target is followed by a transition radiation detector
(TRD) \cite{trd}, a preshower (PRS) and an electromagnetic
calorimeter (ECAL) \cite{ecal}. A hadron calorimeter and two muon
stations are located just after the magnet coil. The neutrino interaction
trigger \cite{trigger} consists of a coincidence between two planes of
counters located downstream of the active target, in the absence of a signal from
a large area system of veto counters upstream of the NOMAD detector.\\
\noindent The CERN-SPS wide band neutrino beam is produced by $450~GeV/c$ 
protons incident on a beryllium target. Neutrinos are produced in the 
decay of secondary pions and kaons in a $290~m$ long decay tunnel at 
an average distance of $625~m$ from the detector. The relative beam 
composition is predicted to be 
$ \nu_{\mu} : \bar\nu_{\mu} : \nu_{e}: \bar\nu_{e}= 1.00:0.0612:0.0094:0.0024$ 
with average energies of  23.5, 19.2, 37.1 and 31.3 $GeV$ 
respectively \cite{gianmaria}.
 The average energy of muon neutrinos interacting in the 
apparatus is about 41 $GeV$.\\
\noindent Referring to Fig.~\ref{detector} the coordinate system adopted
for NOMAD has the $X$-axis into the plane of the figure, the $Y$-axis 
upwards and the $Z$-axis horizontal, approximately
along the direction of the neutrino beam, which points upward, at an angle of
$2.4^{\circ}$ with respect to the $Z$-axis. Its origin is in the center of 
the front face of the first DC along the beam.

\section{The data sample}
\label{data}

\subsection{Event selection}
\label{evsel}
This study is based on the NOMAD full data sample collected between 
1995 and 1998. 
In this analysis the event selection requires a primary vertex with at 
least 2 tracks inside the fiducial area defined by 
$\vert X \vert <130~cm$ and $ -125< Y <135~cm$. Along the $Z$ direction
a cut $Z_{VMIN} < Z < 400~cm$ is imposed. $Z_{VMIN}$ is $5~cm$ for the  1995
and 1996 data and is $35~cm$ for the data collected in 1997 and 1998. 
In the second period
the first DC module was removed to install NOMAD--STAR \cite{star}
and the fiducial volume was slightly reduced. \\
To select a \numucc event a negative $\mu$ attached to the primary vertex, 
identified by the muon stations and having a momentum of at least $3~GeV/c$,  
was required. Under these conditions the total \numucc sample consists of 944019 events.

\subsection{Track selection}
\label{tracksel}
Only tracks attached to the primary vertex 
are used in the search for backward particles. The track is required to 
have at least 8 DC hits,
the distance of the first hit from the primary vertex
to be less than 15 cm, and the relative error on the reconstructed 
momentum to be less than 0.3.
For the \bp search, since the identification method makes use of the
momentum--range relation, tracks are required to range out in the DC volume.
To avoid escaping tracks we require in addition the 
last hit of the track to be in a reduced fiducial volume defined by
$\vert X \vert <100~cm $, $\vert Y \vert < 100~cm $ and $Z_{MIN} < Z < 375~cm$, 
where $Z_{MIN}=30~cm$ for the 1995 and 1996 data sets and  $Z_{MIN}=60~cm$ for the rest. 
Finally no secondary vertex must be found close to the last hit of the track.

\subsection{MC sample}
\label{mc}
In order to estimate proton and pion reconstruction efficiencies and 
background we use a large sample of \numucc Monte Carlo (MC) events.
These events are generated using a modified version of LEPTO 6.1 \cite{lepto}
and JETSET 7.4 \cite{jetset}.
The target nucleon motion is simulated using the Fermi momentum distribution 
proposed by Bodek and Ritchie \cite{bodek}.
The generated events undergo a full detector simulation based on 
GEANT \cite{geant} and are subsequently reconstructed.
MC events and tracks are selected in the same way as are experimental
data yielding a final MC sample consisting of $\simeq 2.5 \times 10^6$ events.
The MC data used for this analysis do not include the simulation of 
``nuclear effects'' such as the intranuclear cascade or correlations
\cite{nom2}.
When needed, specific correction procedures to the MC events 
are applied using the experimental data distributions (see Sec.~\ref{datacorr}).

\subsection{Backward proton identification}
\label{bp_id}

\begin{figure}[htb]
\begin{center}
 \vskip -0.5cm
 \epsfig{file=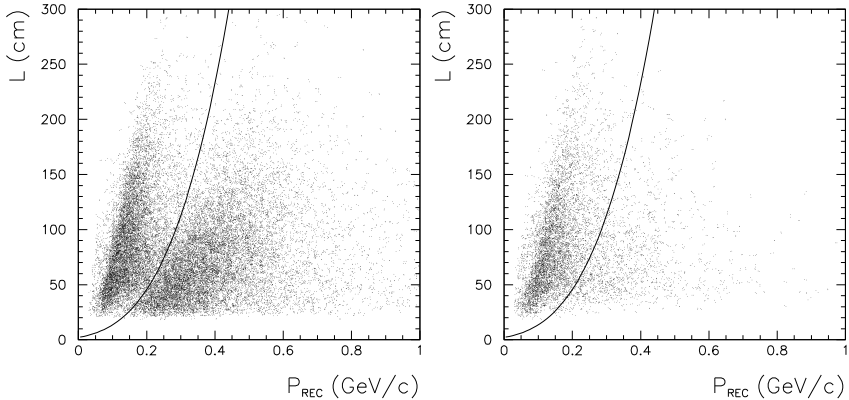,width=120mm}
\end{center}
\vskip -.2cm
\caption{{\it Distributions of length vs. momentum for positive 
(left) and negative (right) backward going tracks selected as described
in the text. The line indicates the position of the cut (Eq.~\ref{lcut}). }}
\label{lenvsp}
\end{figure}

Fig.~\ref{lenvsp} shows the experimental distributions 
of length vs. momentum for positive (left) and negative (right) 
tracks, going backward with respect to the beam direction, and satisfying
our selection criteria. 

\begin{figure}[htb]
\begin{center}
 \vskip -0.5cm
 \epsfig{file=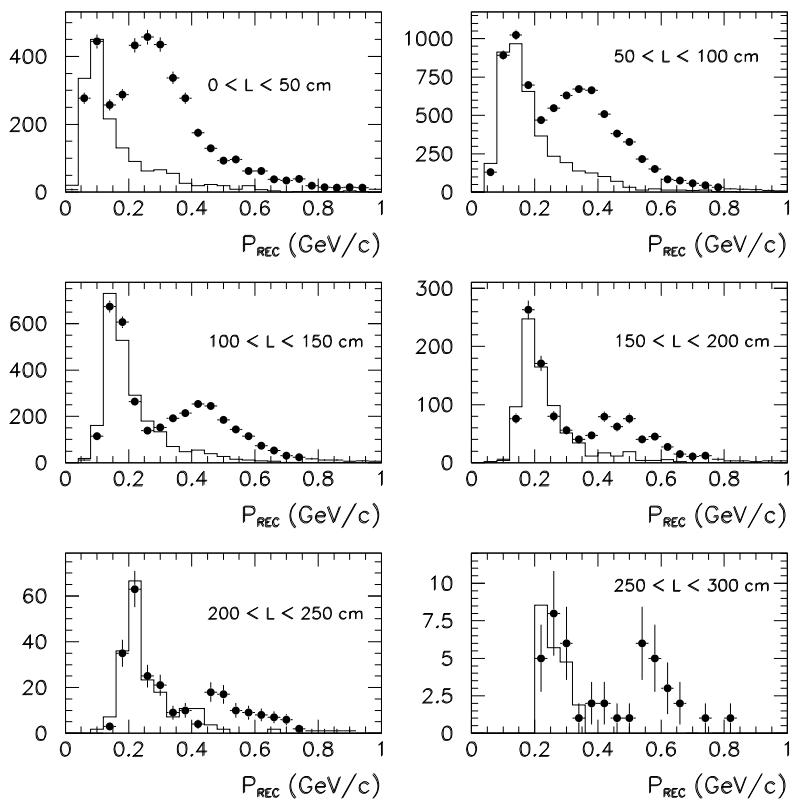,width=120mm}
\end{center}
\vskip -.5cm
\caption{{\it Reconstructed momentum distributions for positive 
(dots) and negative (solid line) backward going tracks for different intervals
of the track length. The heights of the negative distributions have been 
normalized to the positive pion peaks.  }}
\label{two_peak}
\end{figure}

\noindent Two distinct populations are clearly visible on the positive sample plot.
Protons, having a shorter range than $\pi^+$, tend to accumulate in the
lower right part of the plot while the $\pi^+$'s tend to populate 
the left--hand side. Comparing the two plots we see that the lower right
part of the negative sample is much less populated than the corresponding 
region of the positive one, these tracks being mainly $\pi^-$. 
The separation between protons and pions
is also visible in Fig.~\ref{two_peak} where the momentum distribution of
positive (dots) and negative (solid line) backward tracks is shown for 
different intervals of track length. 
In this figure the heights of the negative distributions have been 
normalized to the positive pion peaks.\\
\noindent We identify as a proton any positive 
backward going track which passes our selection cuts and has length $L$:
\be
L \le 2000~(P_{REC}+0.150)^{3.6}~~cm
\label{lcut}
\ee

\noindent where $P_{REC}$ is the reconstructed track momentum in $GeV/c$.
The cut position was optimized by using the MC results of forward tracks
and is shown in Fig.~\ref{lenvsp}.

\subsection{Data corrections for the \bp analysis}
\label{datacorr}
\begin{figure}[htb]
\begin{center}
 \epsfig{file=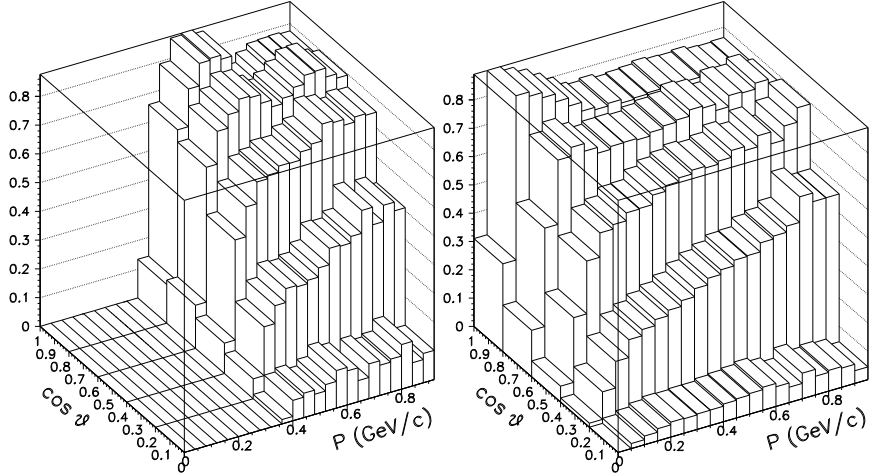,width=120mm}
\end{center}
\vskip -.5cm
\caption{{\it Reconstruction efficiency (as described in the text) as a 
function of momentum and angle between the 
track and the nominal beam direction. 
The plot on the left is for protons, the one on the right for $\pi^{-}$. }}
\label{effric}
\end{figure}

The reconstruction procedure used the energy loss of a pion for the track
fit. Instead of refitting proton candidate tracks with the correct energy 
loss, an empirical correction was applied to the reconstructed proton momentum
$P_{REC}$. This correction, which was obtained from MC events, amounts to
$+180~MeV/c$ at $P_{REC}=200~MeV/c$ and becomes negligible above $500~MeV/c$.
This procedure gives the true proton momentum $P$.\\
The raw number $N_{p}^{raw}$ of identified protons in a given momentum
bin must be corrected  for reconstruction efficiency $\epsilon_{REC}$,
stopping efficiency $\epsilon_{STOP}$, identification efficiency $\epsilon_{ID}$
and pion contamination in the data $\pi^{+~data}_{cont}$. All these
corrections will be described below. They are functions of either the
reconstructed or the true proton momentum.
Since the MC used for this analysis does not properly account for 
\bp production, some of these efficiencies can only be obtained 
by correcting the MC efficiencies with the data, or directly from data
themselves. The ``true'' number $N_{p}$ of protons is then:
\begin{equation}
N_{p} ~=~ N_{p}^{raw} ~~~~{1 \over \epsilon_{REC}}~~~~{1 \over \epsilon_{STOP}} 
~~~~{1 \over \epsilon_{ID} }~~~(1 - \pi^{+~data}_{cont} )
\label{p_corr}
\end{equation}
\noindent with an average overall correction factor of 3.9~.
\vskip 0.1 cm
{\em a) Reconstruction efficiency $\epsilon_{REC}$ }\\
This is the ratio between the number of 
reconstructed and generated protons in a specific bin of the true 
momentum $P$ and angle $\theta$ (measured with respect 
to the beam direction). Since this quantity depends only on the detector
geometry and composition, we assume that the predictions obtained using
forward going protons are applicable to the backward ones with the
replacement $\cos\theta \rightarrow -\cos\theta$. An average value of 0.48
is obtained for $\epsilon_{REC}$, the distribution of which is 
shown on the left in Fig.~\ref{effric}.
\vskip 0.05 cm
{\em b) Stopping efficiency $\epsilon_{STOP}$}\\
This is the ratio between the number of backward protons fully 
contained in the target ($Bp_{STOP}$) and all the reconstructed ones (\bp). 
Its average value is 0.42 in the momentum range used in this analysis.\\
\noindent The protons are extracted from the sample of positive backward tracks by 
subtracting the backward $\pi^+$ content. 
The stopping efficiency $\epsilon_{STOP}$ is computed from the following 
quantities:
\bmath
\epsilon_{STOP} = {Bp_{STOP} \over Bp} = 
{B^+_{STOP}~-~~B\pi^{+}_{STOP} \over B^+~-~B\pi^{+}}
\emath
\noindent where $B^+$ ($B^+_{STOP}$) refer to 
 backward (backward contained) positive tracks. $B\pi^{+}$ and $B\pi^{+}_{STOP}$ 
 are the analogous quantities for backward positive pions.
They are evaluated from the data as follows.\\
In the NOMAD isoscalar target the length distributions of $\pi^+$ and $\pi^-$,
in each momentum bin, are the same apart from a scale factor. Therefore
the number of backward $\pi^{+}$ can be computed by the number of backward
$\pi^{-}$ once the relative population of $B\pi^{+}$ and $B\pi^{-}$
is known. The ratio of the two populations is measured from a clean sample
of pions obtained by selecting tracks with a sufficiently longer length than the
corresponding proton range.
This ratio has comparable values for contained and not contained
pions. It is almost constant in the momentum range of interest with
a weighted average of $R~=~1.82\pm0.05$.\\
The pion content in the backward sample is therefore:\\
\bmath
B\pi^+ ~=~ R \times B\pi^{-}~~~;~~~B\pi^{+}_{STOP} ~=~ R \times B\pi^{-}_{STOP}
\emath
\noindent where the negative pion sample ($B\pi^{-}$, $B\pi^{-}_{STOP}$) is
taken to be the sample of negative (negative contained) backward tracks.
This relies on the fact that the MC simulation shows that the $e^{\pm}$ 
contamination in the backward region is at the level of a few percent above 
$\approx 200~MeV/c$.
\vskip 0.05 cm
{\em c) Identification efficiency $\epsilon_{ID}$ }\\
This is the fraction of \bp with track length smaller than the length 
cut. Above $P_{REC} \simeq 300~MeV/c$ this quantity is $1$.
The results of forward going protons are applicable also in this case.
\vskip 0.05 cm
{\em d)  Pion contamination $\pi^{+~data}_{cont}$}\\ 
This is the fraction of  $\pi^+$'s in the sample of identified protons,
which amounts to  $\approx 8\% $ above $P_{REC} \simeq 250~MeV/c$.
To estimate this contamination in the data, the MC prediction ($\pi^{+~MC}_{cont}$) 
for forward particles has been corrected using the backward data 
distributions as follows. 
For each momentum bin the length distributions of protons
in the data and MC can differ in normalization but not in shape 
since the track length depends on the mechanism of energy loss which is well 
reproduced by the MC. The same applies to $ \pi^{+}$.
The length cut applied on the momentum--length plane (Eq.~\ref{lcut})
defines in the data and MC two populations: the $\pi$--like tracks, (the tracks
above the cut) and the $p$--like tracks (those below the cut). 
To account for the difference in the populations 
of protons and  pions in the data and MC the quantity $ \pi^{+~MC}_{cont}$
has been weighted by the double ratio of the $\pi$--like tracks (data/MC)
and $p$--like tracks (data/MC):
\bmath
 \pi^{+~data}_{cont}~=~{ N_{\pi^{+}}^{data} \over N_{p}^{data} }~=~
 { N_{\pi-like}^{data} / N_{\pi-like}^{MC} \over 
   N_{p-like}^{data} / N_{p-like}^{MC} }~~ \pi^{+~MC}_{cont} 
\emath
\noindent The above correction is evaluated as a function of the
reconstructed momentum.\\

\subsection{Backward $\pi^-$ identification}
\label{bim_id}

\noindent As already pointed out the $e^-$ component is negligible in 
the negative backward track sample. 
We therefore assume that any negatively, backward going, 
charged track with $P > 0.2~GeV/c$ is a $\pi^-$. 
The smallness of the contamination and the fact that it is not necessary 
to look for stopping tracks to identify the \bpim reduce the number of 
corrections to be applied to the data  to only one, the reconstruction 
efficiency $\epsilon_{REC}^{~\pi^{-}}$.For this quantity we use the MC 
predictions obtained for forward going $\pi^-$.\\
The true number of negative pions is then obtained from the raw number of 
identified negative pions $N_{\pi^-}^{raw}$ in a given momentum bin 
according to:

\begin{equation}
N_{\pi^{-}} ~=~ N_{\pi^-}^{raw} ~~~~{1 \over \epsilon_{REC}^{~\pi^{-}} }
\label{pim_corr}
\end{equation}

\noindent In this case the average value of the correction is 2.

\section{Backward $p$ and $\pi^-$ invariant momentum distributions}
\label{invslope}

\begin{figure}[htb]
\begin{center}
   \epsfig{file=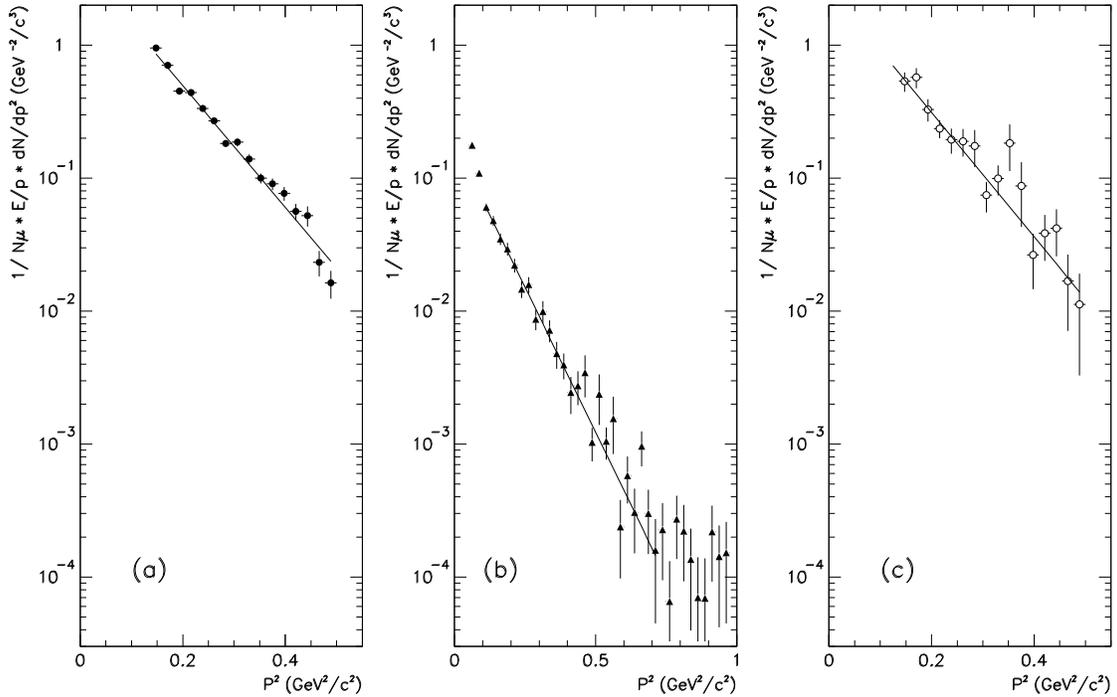,width=160mm}
\end{center}
\vskip -.5cm
\caption{{\it Invariant momentum distributions for backward going protons 
(a), $\pi^-$ (b) in $\nu_\mu CC$ events and protons in $\bar\nu_\mu CC$ events (c)}}
\label{slopes}
\end{figure}

\noindent The inclusive spectrum of backward particles is typically represented 
using the normalized invariant cross section $ (1/\sigma_{TOT})~E~d^3\sigma/dP^3$,
where $E$ is the energy of the backward going particle. The invariant
cross section is usually \cite{matsinos} parametrized by an exponential form as:\\

\be
{1 \over N_{ev} }~{E \over P}~{dN \over dP^{2}}~=~C~e^{-BP^{2}}
\label{bparam}
\ee

\noindent where $N_{ev}$ is the total number of events, 
$C$ is a constant and $B$ is the slope parameter. In previous experiments 
$B$ has been found to be almost independent of the projectile type and momentum, 
and of the atomic number of the target.
This behaviour has been termed ``{\em nuclear scaling}'' \cite{fredriksson}.\\
The inclusive spectrum of \bp and \bpim is shown in Fig.~\ref{slopes}a and 
\ref{slopes}b respectively, together with the exponential fit. 
The values of the measured slope parameter $B$ are reported in 
Table~\ref{table:slope_table}. The first and the second errors are statistical 
and systematical, respectively.

\begin{figure}[htb]
\begin{center}
   \epsfig{file=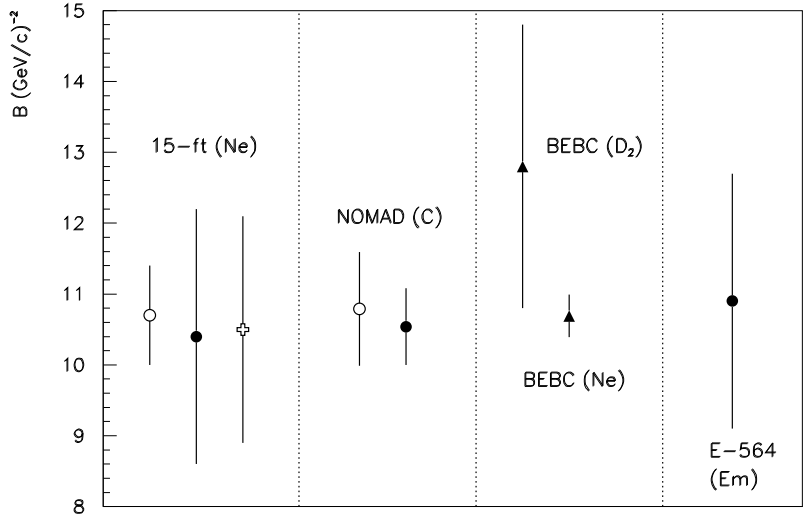,width=120mm}
\end{center}
\vskip -0.5cm
\caption{{\it 
The value of the slope parameter $B$ obtained in neutrino experiments. 
For NOMAD the statistical and systematical error have been added in quadrature. 
The full circles are the results from $\nu_{\mu}~CC$, 
the open circles from \anumucc and  the cross is for \numunc interactions. 
For BEBC $\nu$ and $\bar\nu$ results have been combined.}} 
\label{bnu}
\end{figure}

\begin{table}[ht]
\centering
\begin{tabular}{llcll}
\hline
                 &          & $\Delta P~(GeV/c)$ & $~~~B (c^{2}/GeV^{2})$ & $~~~C (c^3/GeV^2)$  \\ \hline
Backward $p$     & \numucc  & $0.37 \div 0.70$   & $10.54\pm0.20\pm0.5$ & $4.08\pm0.19\pm0.5$ \\ 
                 & \anumucc & $0.37 \div 0.70$   & $10.79\pm0.78$       & $2.71\pm0.54$       \\ 
Backward $\pi^-$ & \numucc  & $0.32 \div 0.85$   & $10.03\pm0.28\pm0.3$ & $0.17\pm0.01\pm0.02$\\ \hline \\
\end{tabular}
\caption{{\it Fit ranges and values of the slope parameter $B$, for backward protons and
$\pi^{-}$, as obtained from the exponential fit to the invariant cross section.
The first and the second errors are statistical and systematical, respectively.}}
\label{table:slope_table}
\end{table}
\noindent The systematic uncertainty was estimated by slightly changing
the values of all the cuts used, by varying by a small amount the correction functions
and also by changing the size of the fiducial volume. The values of the slope parameters
 measured in this experiment are found to be compatible with the
results obtained in other neutrino (see Fig.~\ref{bnu} for the \bp case) and
hadronic experiments.
\noindent The invariant cross section for \bp is larger than the one for
\bpim by about one order of magnitude but the values of the slopes are
similar. The kinematic ranges of the two invariant distributions are also different.
To be identified \bp have to stop inside the target volume; the rather low
density of the NOMAD target (see Sec.~\ref{nomad}) restricts to $\approx 0.7~GeV/c~$
the maximum useful momentum value.\\
\noindent The invariant cross section and slope parameter for \bp in \anumucc events 
are given in Fig.~\ref{slopes}c and Table~\ref{table:slope_table} respectively.
During normal operations \anumucc events were also collected due to the small
$\bar\nu_{\mu}$ component of the dominantly $\nu_{\mu}$ beam \cite{nom1}.
A dedicated $\bar\nu$ run yielded an additional sample of \anumucc events
included for this analysis.\\
Antineutrino events are selected under the assumptions that the efficiencies and 
pion contamination are the same as those used for the neutrino events,
but requiring a positively charged muon instead of a negative one.
The final sample consists of $61134$ events containing 1764~\bp.

\subsection{Energy dependence of the slope parameter}

\begin{figure}[htb]
\begin{center}
   \epsfig{file=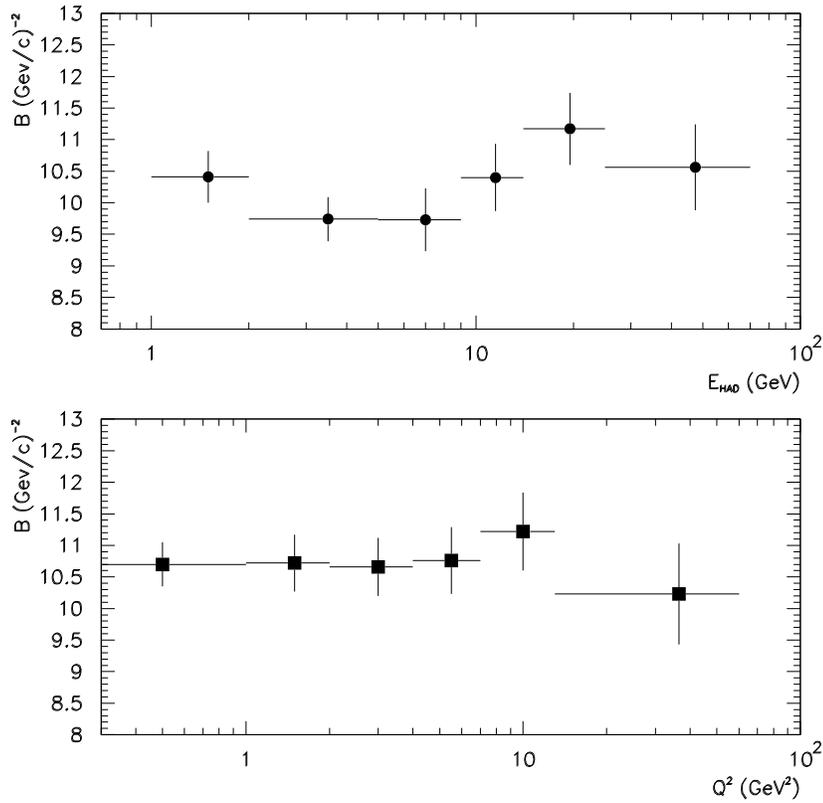,width=120mm}
\end{center}
\vskip -.5cm
\caption{{\it 
The value of the slope parameter $B$ shown as a function  
of the hadronic energy $E_{HAD}$ (top) and of $Q^2$ (bottom). }}
\label{behad}
\end{figure}

\noindent
In Fig.~\ref{behad} we show the slope parameter $B$ for protons as a function
 of the visible hadronic energy $E_{HAD}$ and the $Q^2$ of the event.
The visible hadronic energy is defined as: 
\bmath
E_{HAD}= \sum E_{c} + \sum E_{n} 
\emath
\noindent where $\sum E_{c}$ is the sum of the energies of reconstructed charged tracks
(assuming the mass of the pion if the particle type is not explicitly identified).
The sum includes all the charged tracks attached to the
primary vertex by the reconstruction program, those belonging to a 
pointing $V^{0}$, the ``close hangers'' i.e. 
isolated tracks not associated to the primary nor to a secondary vertex
and having their first hit in a box around the primary vertex of size
$\vert X \vert$, $\vert Y \vert <10~$ cm and $-5<Z<20~$ cm, and the ``pointing
hangers'' i.e. the hangers which have an impact parameter of less than 10 cm
when linearly extrapolated back to the z-plane of the primary vertex.
$\sum E_{n}$ includes tracks associated to secondary vertices corresponding 
to interactions of neutral particles and the energy of neutral particles 
reconstructed in the ECAL. The reconstructed neutrino energy $E_{vis}$
is taken to be the sum of the muon energy $E_{\mu}$ and of $E_{HAD}$.
The square of the four--momentum transfer $Q^2$ is 
$Q^2 = 4~E_{vis}E_{\mu}sin^2\theta/2$, where $\theta$ is the muon angle 
with respect to the neutrino direction.
\noindent No significant dependence of the slope $B$ on either quantity is observed, 
in agreement with the expectations of ``nuclear scaling'' as observed in hadronic 
beam experiments. The range covered by NOMAD is similar to the one covered by different
experiments with hadronic beams.

\subsection{Angular dependence of the slope parameter}
\begin{figure}[htb]
\begin{center}
   \epsfig{file=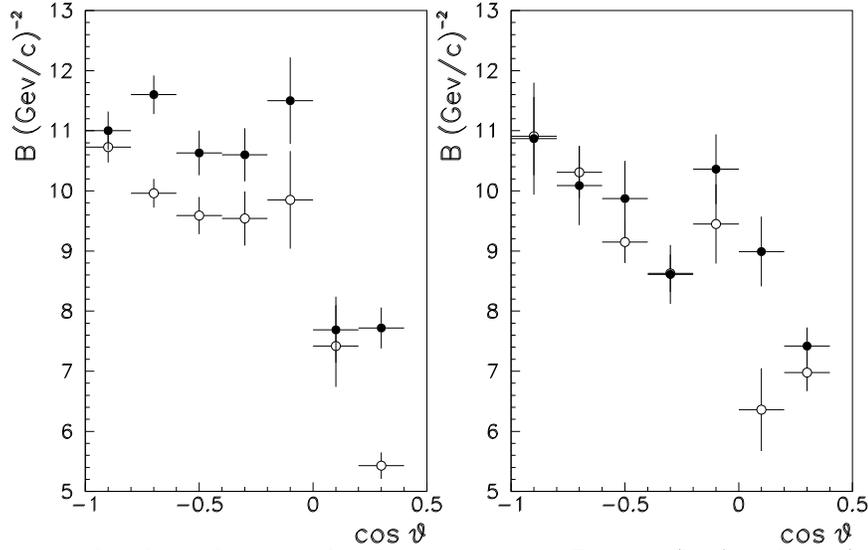,width=120mm}
\end{center}
\vskip -.5cm
\caption{{\it 
Angular dependence of the slope parameter $B$ for $p$ (left) and $\pi^{-}$ (right).
The value of $B$ is shown as a function of the particle angle with respect to the 
hadronic jet direction (full circles) and with respect to the beam 
direction (open circles). }}
\label{bcosthe}
\end{figure}

The angular dependence of the slope parameter $B$ is usually given with respect
to the direction of the exchanged boson, approximated by the hadronic jet direction
\cite{matsinos}. This allows a comparison with the results obtained in 
hadron beam experiments. The angular dependence of $B$ 
for \bp and \bpim is shown in Fig.~\ref{bcosthe} as a function of the angles with 
respect to  the beam and to the hadronic jet direction. 
Although these two angles are highly correlated, there are small differences in 
the slope values. If the beam direction is used $B$ is systematically smaller by 
$\approx 0.5\div1 (GeV/c)^{-2}$, in the \bp case, while for \bpim the difference is less 
pronounced. As can be seen from Fig.~\ref{bcosthe}, 
larger values of $B$ are preferred at increasingly backward directions.
This behaviour has already been observed in neutrino \cite{matsinos}, 
photon \cite{alanakyan} and hadron \cite{bayukov,komarov,yulda}
nucleus experiments.

\section{Kinematical properties of events with a backward particle}
\label{kine}

Table~\ref{table:kine_table} shows a comparison of the average values of some
kinematical quantities for \numucc events with and without \bp or $B\pi^{-}$.
$<C>$ and $<N^{\pm}>$ are the average event charge and charged multiplicity 
($\mu^{-}$ included), $\nu = E_{vis}-E_{\mu}$, $x = Q^{2}/2M\nu$ is the 
Bjorken scaling variable and $W^{2} = M^{2}-Q^{2}+2M\nu$ is the square of the 
invariant mass of the hadronic system, $M$ being the nucleon mass.\\
\begin{table}[htbp]
\centering
\begin{tabular}{crrr}
\hline
             & CC~(no~$Bp$, no~$B\pi^-$) & CC~($Bp$) & CC~($B\pi^-$) \\ \hline
$<C> $       &  0.46              &  2.11$\pm$0.02    &   0.08$\pm$0.02 \\ 
$<N^{\pm}>$  &  5.13              &  6.63$\pm$0.03    &   7.28$\pm$0.03 \\ 
$<E_{vis}>$  & 40.75              & 38.9$\pm$0.4      &  39.5$\pm$0.5   \\ 
$<E_{HAD}>$  & 12.32              & 11.4$\pm$0.2      &  11.7$\pm$0.2   \\ 
$<Q^{2}>  $  &  7.07              &  5.5$\pm$0.1      &   5.9$\pm$0.1   \\ 
$<x>      $  &  0.32              &  0.26$\pm$0.002   &   0.28$\pm$0.003\\ 
$<W^{2}>  $  & 17.19              & 16.4$\pm$0.2      &  17.5$\pm$0.3   \\ \hline
\end{tabular}
\caption{\it {Comparison of average values of some kinematical variables in events
with and without an identified \bp or \bpim. The values of $<C>$ and of 
$<N^{\pm}>$ are corrected for the track reconstruction efficiency.
Only statistical errors are shown. They are negligible for the first data column.}}
\label{table:kine_table}
\end{table}
\noindent For events with a backward particle the average values of the
charged multiplicity are consistent with the presence of 
extra tracks as expected in both models. 
The average value of the event charge in the \bp sample ($<C>=2.11$) is 
larger than in the no~$Bp$, no~$B\pi^-$
samples as measured in the data ($<C>=0.46$). It is also larger than in the total 
MC sample where no nuclear effects are present ($<C>=0.31$). 
This observed increase is not entirely due to the bias introduced by demanding 
a positive particle, since an average value of $<C>=1.21$ is obtained when 
requiring a backward positive track in the MC events.
This can be expected from both mechanisms outlined in the Introduction.
In the collisions of secondary hadrons with nucleons inside the
nucleus the total event charge will increase by one unit in interactions 
off protons while no extra charge will be produced when scattering on neutrons.
Also in the framework of the short range correlation model (in its simplest form of
two--nucleon correlations) the mechanism of pair breaking predicts extra
tracks in the final state. The mechanism is of course symmetric for $p$ and $n$
but backward going neutrons are not detected.\\
\noindent The events with a \bpim show an average charge smaller by $\simeq 0.4$
units with respect to the value where no \bpim is present.
Here again this effect cannot be entirely interpreted as the result of the
bias associated with the requirement of at least one negative backward track, 
since the MC in this case predicts $<C>=-0.41$.
Therefore also in the sample of events with a \bpim  an increase in the 
average value of the event charge is observed.\\
From Table~\ref{table:kine_table} it appears that the average values of the 
kinematical variables, other than $<C>$ and $<N^{\pm}>$, for events with either 
a \bp or a \bpim are systematically lower than for events without identified 
backward particles.

\section{Backward particle rates}
\label{bpy}

The \bp rate has been compared with the results
of the other $\nu$--nucleus experiments in order to study a
possible atomic number dependence. In NOMAD, where \bp are identified in the momentum
interval from $370 MeV/c$ to $700 MeV/c$ we obtain, after correction,  the 
number of events reported in Table~\ref{table:nombpy_table}.

\begin{table}[ht]
\centering
\begin{tabular}{crr}
\hline
 N. of \bp or \bpim & N. of \bp ev. & N. of \bpim ev.\\ \hline
  0              & 904212        & 939617\\ 
  1              &  37634        &   4238\\ 
  2              &   2168        &    164\\ 
  3              &      5        &      0\\ \hline
\end{tabular}
\caption{\it {The number of events, after correction for efficiencies,
 as a function of the multiplicity of backward protons 
 in the momentum interval $370\div700 MeV/c$, and of backward negative pions
 in the momentum interval $350\div800 MeV/c$.}}
\label{table:nombpy_table}
\end{table}
\noindent These figures correspond to an average number of 
\bp per event, $<N_{Bp}> = (44.5 \pm 0.5)\times 10^{-3}$,
where the error is statistical only.
\begin{table}[htbp]
\centering
\begin{tabular}{lrrrrr}
\hline
                     &       &                &             &                &                  \\
Experiment           & $<A>$ & $\Delta~P$~~~  & $<E_{\nu}>$ & $ <N_{Bp}> $   & $ <N_{B\pi^-}> $ \\ 
                     &       & (GeV/c)        & (GeV)       & ($\times 10^{-3}$)~~ & ($\times 10^{-3}$)~~\\ \hline
BEBC $D_2$ \cite{d2} & 2     & 0.35$\div$0.80 & 50          & 6.3   $\pm$ 0.7 & 1.0 $\pm$ 0.3 \\ 
NOMAD                & 12    & 0.37$\div$0.70 & 41          & 44.5  $\pm$ 0.5 & 4.8 $\pm$ 0.1 \\ 
BEBC $Ne$ \cite{matsinos} & 20 & 0.35$\div$0.80 & 50        & 97.7  $\pm$ 3.7 & 6.8 $\pm$ 0.9 \\ 
15foot \cite{berge}  & 20    & 0.20$\div$0.80 & 50          & 137.5 $\pm$ 15  & \\ 
SKAT \cite{skat}     & 52    & 0.32$\div$0.70 & 10          & 185.8 $\pm$ 10  & \\ 
E--564 \cite{dayon}  & 80    & 0.30$\div$0.80 & 60          & 331.1 $\pm$ 47  & \\ \hline \\
\end{tabular}
\caption{\it {Comparison of \bp and \bpim rates for neutrino-nucleus 
experiments. The average mass number $<A>$, the momentum interval 
$\Delta P$ used in the analysis, the average event energy $<E_{\nu}>$ and the  
\bp and \bpim rates are listed for each experiment.
The errors are purely statistical. }}
\label{table:bpy_other}
\end{table}
\noindent The backward proton yield obtained in other $\nu$--nucleus experiments is
shown in Table~\ref{table:bpy_other}.
The values, which were extracted from the 
original references, are not directly comparable because
of the different momentum intervals used in the different experiments.
\noindent In order to study the $A$ dependence the \bp yields measured in 
NOMAD and E--564 \cite{dayon} were extrapolated to the interval from 350 to 800 MeV/c  
assuming the dependence given in Eq.~\ref{bparam} with the measured slopes.\\
The extrapolation for NOMAD gives:

\begin{displaymath}
<~N_{Bp}>_{350\div800 MeV/c}~ = 
 \left[ \rule{0mm}{5mm} ~52.8 \pm 0.6 (stat.) \pm 7 (syst.)~ \right] \times 10^{-3} 
\end{displaymath}

\begin{figure}[htb]
\begin{center}
   \epsfig{file=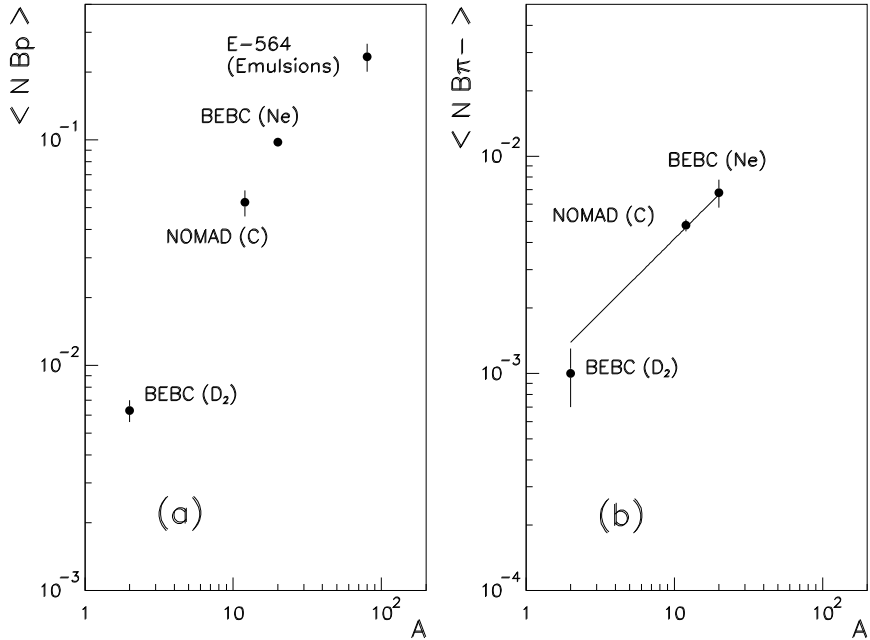,width=130mm}
\end{center}
\vskip -.5cm
\caption{{\it The average number of \bp per event (a) and of \bpim per event (b)
in the momentum range from  $350\div800 MeV/c$ 
in neutrino experiments as a  function of the mass number $A$.
The line shown in (b) is the result of the fit described in the text. }}
\label{bpyield}
\end{figure}

For E--564 we obtain $<~N_{Bp}>=(234\pm33) \times 10^{-3}$.\\
The $A$ dependence for experiments most directly comparable to NOMAD
is shown in Fig.~\ref{bpyield}a. In the range $A=20\div80$
it has been parametrized as $ <N_{BP}>\propto A^{\alpha}$, 
where $\alpha=0.68\pm0.12$ \cite{dayon}.
The same parametrization with a similar power law was found to describe
\bp production in $\pi^{+}$ and $K^{+}$ collisions with $Al$ and $Au$ 
nuclei at 250 GeV/c \cite{na22}.
It is evident from Fig.~\ref{bpyield}a that this simple power law does not
describe the NOMAD data taken at a lower $<A>$.
\noindent The \bpim rate was directly measured in the same momentum range
used for the \bp analysis. Its average value is found to be: 
\begin{displaymath}
<~N_{B\pi^-}>_{350\div800 MeV/c}~ = 
 \left[ \rule{0mm}{4mm}  ~4.8 \pm 0.1 (stat.) \pm 0.3 (syst.) \right] \times 10^{-3} 
\end{displaymath}
\noindent In this case a good fit to the two BEBC points and to the NOMAD value 
can be obtained using the form $A^{\alpha}$ giving $\alpha=0.83\pm0.25$, 
as shown in Fig.~\ref{bpyield}b.

\section{Comparison of data with models}
\label{data&model}
\subsection{$E_{HAD}$ and $Q^2$ dependence of \bp and \bpim rates}

\begin{figure}[htb]
\vskip -0.5 cm
\begin{center}
   \epsfig{file=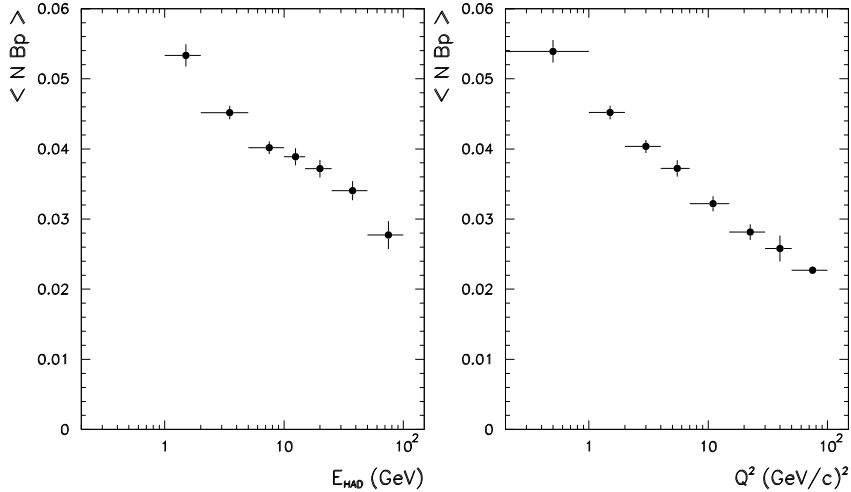,width=120mm}
\end{center}
\vskip -.3cm
\caption{{\it The average number of \bp ($ 370<P<700 MeV/c $) 
per event as a function of the hadronic energy $E_{HAD}$ (left) and of $Q^2$ (right).}}
\label{bpy_ehad}
\end{figure}

\begin{figure}[htb]
\begin{center}
   \epsfig{file=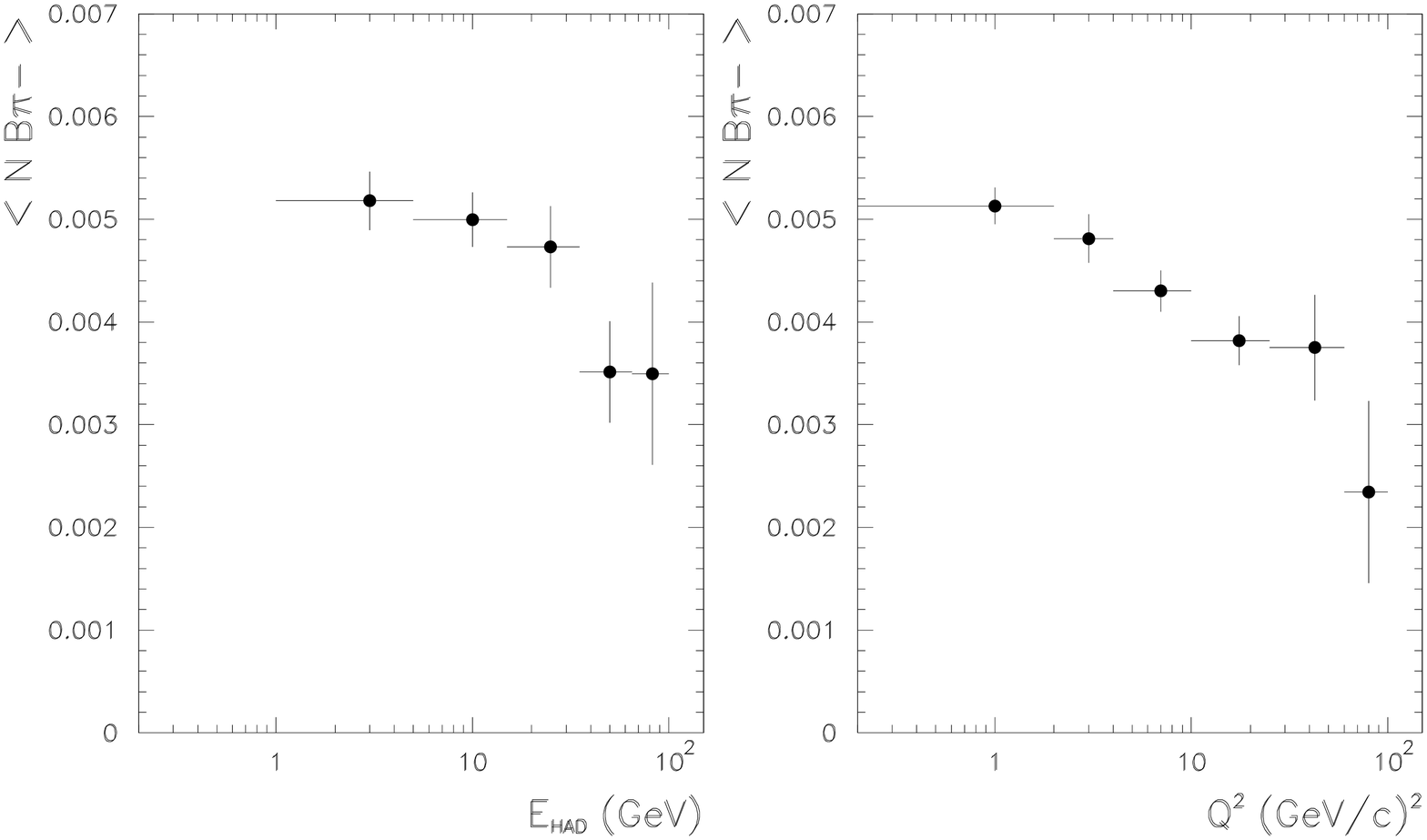,width=120mm}
\end{center}
\vskip -.2cm
\caption{{\it The average number of \bpim ($ 350<P<800 MeV/c $)  
per event as a function of the hadronic energy $E_{HAD}$ (left) and of $Q^2$ (right).}}
\label{bpimy_ehad}
\end{figure}

The \bp and the \bpim rates have been studied as a function of the hadronic energy
$E_{HAD}$ and of $Q^2$. In both cases, shown in Fig.~\ref{bpy_ehad}  
and \ref{bpimy_ehad} a decrease of the yield with increasing 
$E_{HAD}$ and $Q^2$ is observed.
As pointed out in the Introduction, this can be interpreted in terms of the 
``formation zone'' concept. 
The larger $E_{HAD}$ and $Q^2$, the larger is the average energy of the outgoing 
partons therefore resulting in hadrons which have higher probability to
be formed outside the nucleus. 
As a consequence, reinteractions will decrease and so will the slow proton rates.
This $E_{HAD}$ and $Q^2$ dependence is also consistent with the decrease of the 
average values of $E_{HAD}$ and $Q^2$ in events with identified \bp as shown 
 in Table~\ref{table:kine_table}.
In the \bpim case the dependence of the yield on $Q^2$ and on $E_{HAD}$ is 
less pronounced. This can be understood since the \bpim rates are a less 
sensitive probe of nuclear effects because \bpim production on a stationary 
nucleon is kinematically allowed for momenta up to about half the nucleon mass.  
Furthermore \bpim can be produced in the decay of forward going resonances 
or unstable particles. 

\subsection{Multiplicities of slow particles}
\label{slow}
\begin{figure}[htb]
\begin{center}
   \epsfig{file=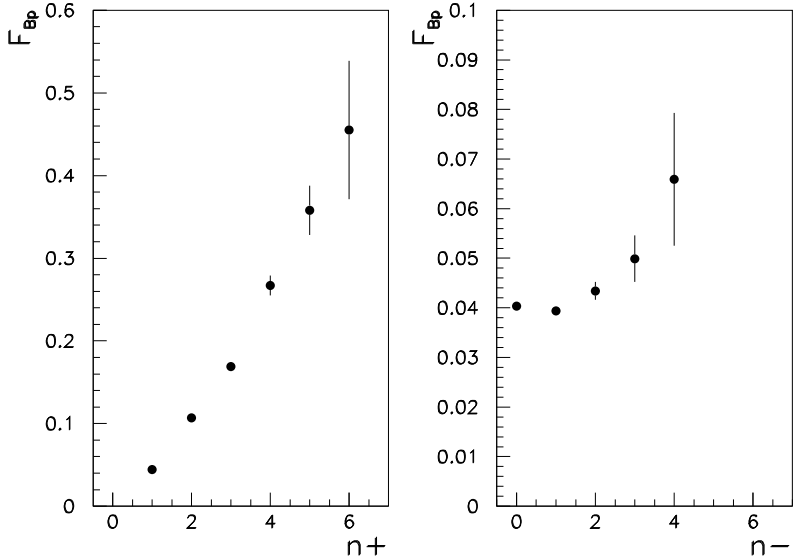,width=120mm}
\end{center}
\vskip -.5cm
\caption{{\it Fraction $F_{Bp}$ of events with at least one \bp
as a function of the multiplicity of low--momentum ($P<700 MeV/c$)
positively (left) or negatively (right) charged hadrons.}}
\label{slow_mult}
\end{figure}

\noindent Since mostly slow hadrons are produced inside the nucleus,
they are predominantly the ones that can scatter off nucleons, producing 
both slow protons and backward particles. Therefore, in the framework of 
the rescattering model, a correlation between the number of emitted slow 
protons and the multiplicity of mesons is expected  \cite{ishii}.\\
The increase in the rate of events with slow protons as a function of
hadron multiplicities was first observed by
the E745 collaboration \cite{kitagaki} in a $\nu$-Freon
experiment. This correlation was used \cite{ishii} to
extract information on the formation zone of secondary hadrons.
This effect was also observed by the BEBC collaboration\cite{guy}.\\
\noindent We have observed this correlation by studying the 
fraction of events with at least one \bp as a function of the 
multiplicity of low momentum ($P<700 MeV/c$) positively or negatively
charged hadrons (Fig.~\ref{slow_mult}). 
In both cases, this fraction of events increases with
increasing multiplicity of low momentum tracks.
However, the positive hadronic multiplicity is biased by the presence
of protons from rescattering itself and therefore it is less representative
of the original multiplicity of positive slow hadrons from the neutrino 
interaction. The true correlation can therefore be better studied as a function
of the multiplicity of negative slow hadrons.

\subsection{Backward $\pi^-$ spectra and the Fermi momentum tail}

As already pointed out in the Introduction, in the
short range correlation models the spectrum of fast backward going
hadrons reflects the tail of the Fermi momentum distribution.
For this study we used \bpim because, as opposed to $Bp$, they do not 
need to range out in the target thus yielding data up to larger momenta.\\
To estimate qualitatively the contribution of this tail we have compared
our \bpim data to the predictions of two different Fermi momentum 
distributions \cite{bodek,benhar}. 
\begin{figure}[htb]
\begin{center}
   \epsfig{file=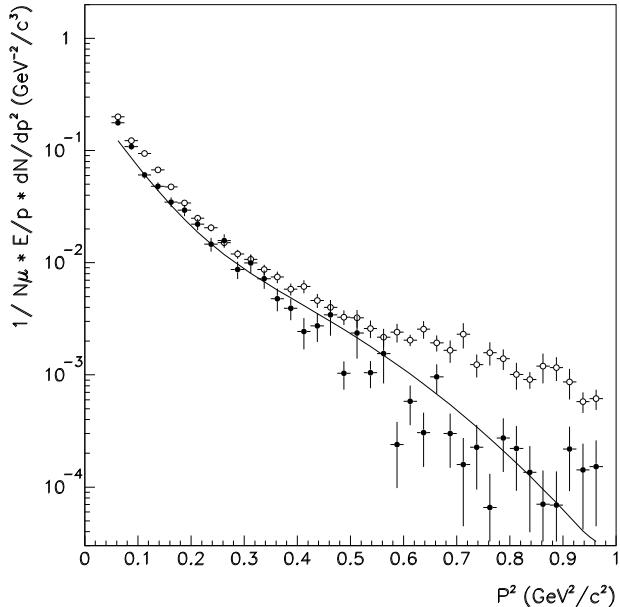,width=90mm}
\end{center}
\vskip -.5cm
\caption{{\it 
Invariant spectrum for \bpim in MC (open circles) and data (full circles). 
The solid line is the curve of Ref.~\cite{benhar} superimposed with an 
arbitrary normalization. }}
\label{mcpion}
\end{figure}
The distribution of Ref.~\cite{bodek} has a long momentum tail up to $4~GeV/c$.
In Ref.~\cite{benhar} the spectrum of backward going hadrons in lepton--nucleon 
scattering is derived from the spectral function $p({\bf k},E)$, 
which is the probability of finding a nucleon with momentum
${\bf k}$ and removal energy $E$. The function, whose integral over $E$ is the 
Fermi distribution, is obtained from non relativistic many body 
theory and a specific procedure has been developed to extrapolate it to 
large values of ${\bf k}$. In this case the Fermi momentum tail is shorter and
ends at about $1~GeV/c$ \cite{benhar_priv}.\\
In Fig.~\ref{mcpion} the invariant spectrum of MC $B\pi^{-}$, simulated 
using the Fermi distribution of Ref.~\cite{bodek}, is shown 
together with the data.
A clear disagreement is visible in the tail of the distributions,
the MC  being much larger than the data above $\simeq 0.5~(GeV/c)^{2}$.
Our results therefore do not support a Fermi momentum distribution
with a long tail as proposed in Ref.~\cite{bodek} while they agree
with the dependence predicted by Ref.~\cite{benhar}, as also 
shown in Fig.~\ref{mcpion}.\\

\subsection{Effects of short range correlations in \bp production}

According to the picture proposed in Ref.~\cite{fs1} \bp production 
is explained as the result of a neutrino interaction within a correlated 
cluster of two or more nucleons. 
This cluster is formed, for a short time, when two nucleons in their motion 
inside the nucleus approach each other so closely as to come under the effect
of the short range component of the nuclear force ($r_c = 0.5 \div 0.7 fm$). 
As a result, the high relative momenta of the correlated nucleon pair
manifest themselves when the backward moving spectator is released in 
the interaction of the incoming virtual $W$ with the forward going nucleon. 
In this model, if the effects of reinteractions are neglected, the released 
backward nucleon can leave the nucleus keeping its original momentum.
These correlated pairs have recently been observed in the 
reaction $e+^{16}O~\rightarrow~e^{'} p p~^{14}C $ at low values of 
the energy transfer $(180\div240 MeV)$ \cite{starink}.\\
To study these correlations it is customary to use the variable 
$\alpha$ defined as:
\be
\alpha=(E~-P_l)/ M
\label{eq_alpha}
\ee
 \noindent where $E$ and $P_l$ are respectively
the energy and  longitudinal momentum of the \bp and 
$M$ is the nucleon mass. For $Bp$, $\alpha > 1$ since $P_l$ is negative.
In this model, due to the target motion a correlation between 
the Bjorken scaling variable $x$ and the variable $\alpha$ is expected. 
In particular the average $x$ for events with a \bp is expected to be smaller
than in events where no \bp is present, as indeed observed in our data 
(see Table~\ref{table:kine_table}). The same correlation is expected to 
hold if the variable $v = xy$ is used, where $y = E_{HAD}/E_{vis}$.
This variable is related to the muon kinematics and  can be written
as : 
\be
v=(E_{\mu}-P_\mu^{l})/M
\label{eq_v}
\ee
\noindent where $E_{\mu}$ and $P_\mu^{l}$ are the muon energy and longitudinal 
momentum. Since, generally, in neutrino experiments the muon kinematic variables 
are well measured, $v$ is better suited than $x$ for this purpose.\\
\noindent The $(v,\alpha)$ correlations were searched for in the data
by calculating for each event $\alpha$ and $v$ as defined in Eq.~\ref{eq_alpha} and
\ref{eq_v}. For each $\alpha$ bin we plot the variable $<v_N>$ defined as:
\be
<v_N>~=~{<v>_{Bp} \over <v>_{no~Bp}}
\ee
\noindent where $<v>_{no~Bp}$ is obtained from the full sample of events 
without a $Bp$. According to Ref.~\cite{fs1} the average values of 
$v$ in events with a $Bp$, $<v>_{Bp}$, is related to the average value of $v$ in events 
where no \bp is present, $<v>_{no~Bp}$, by:
\be
<v>_{Bp}~=~<v>_{no~Bp}~(2-\alpha)
\ee

\noindent More generally for a cluster composed of $\xi$ nucleons
the relation is \cite{multiq}:

\be
<v>_{Bp}~=~<v>_{no~Bp} \left(1 - {\alpha \over \xi} \right)~ {\xi \over {\xi-1}}
\label{v_cluster}
\ee

\begin{figure}[htb]
\begin{center}
   \epsfig{file=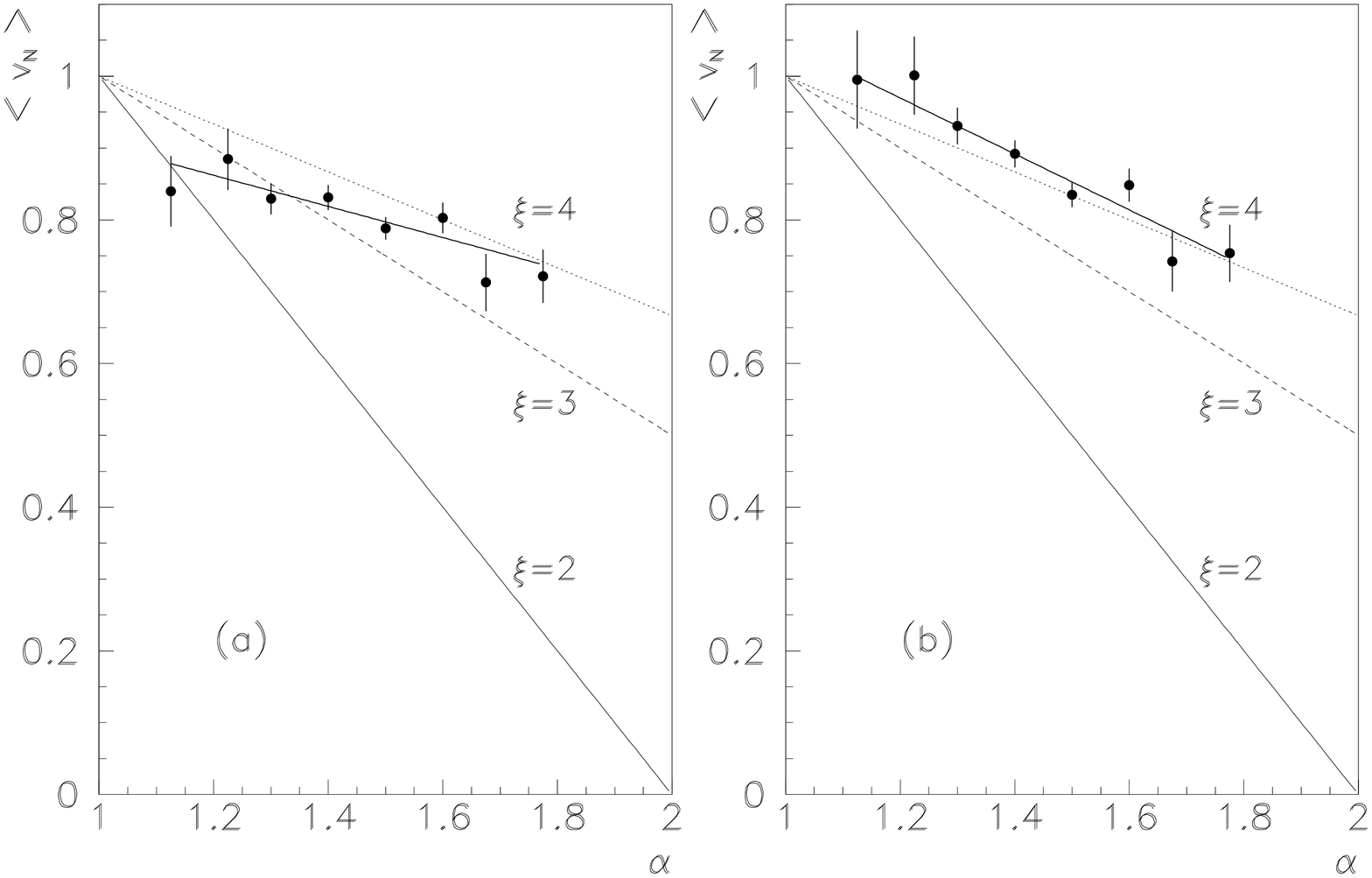,width=120mm}
\end{center}
\vskip -.5cm
\caption{{\it The variable $<v_N>$ plotted as a function of $\alpha$. 
The lines represent the predicted correlation (Eq.~\ref{v_cluster}) 
for a number $\xi$ of nucleons in the cluster equal to 2, 3 and 4. 
In (a) all the \bp events were used; 
in (b) only events having a \bp with $cos{\theta_{j}} < 0$, $\theta_{j}$ 
being the angle of the \bp with respect to the hadronic jet direction.}}
\label{alpha}
\end{figure}

\noindent Fig.~\ref{alpha}a shows  $<v_N>$ as a function of $\alpha$.
The data indicate a slope of $-0.22\pm0.06$ with a $\chi^{2}/ndf=5.3/6$ 
(the $\chi^{2}/ndf$ in the hypothesis of no dependence on $\alpha$ is 19.1/7). 
The two--nucleon correlation mechanism ($\xi=2$, the most probable case when
considering the overlapping probabilities of $\xi$ nucleons in the nucleus)
fails to describe our data. Either higher order structures are playing
a leading role \cite{multiq} or the observed low level of correlation is due
to the presence of reinteraction processes. In this case part of the 
\bp are emitted as a result of reinteractions in the nucleus and are not
related to the target nucleon. The presence of intranuclear cascade processes
could therefore dilute the existing correlation to the observed level.
\noindent To test this hypothesis we tried to reduce the component 
due to rescattering in the selected \bp sample. Having observed the 
correlation existing between the multiplicities of slow tracks
and rescattered protons (Sec.~\ref{slow}) we applied increasingly 
tighter cuts on the number of slow tracks ($P<700~MeV/c$). 
As a consequence of these cuts the degree of correlation between 
$\alpha$ and $v$ increases. 
The fit values of the slopes are reported in Table~\ref{table:alpha}
together with the definitions of the cuts applied.\\
We have also observed a strong correlation between the presence of protons travelling
backward in the lab but forward with respect to the hadronic jet direction, and the 
concentration of events  at small $Q^2$ values and large angles with respect to the beam. 
Since also a small $Q^2$ indicates the presence of rescattering,
the exclusion of these events should highlight the expected correlation.
The resulting slope is $-0.39\pm0.07$ with a $\chi^{2}/ndf=4.9/6$
(see Fig.~\ref{alpha}b).\\
The observed behaviour is consistent with the hypothesis of the correlations
effects being to some degree hidden by the presence of rescattering. 
Reducing the rescattering component these correlations seem to become stronger.

\begin{table}[ht]
\centering
\begin{tabular}{cclc}
\hline 
 ~~$n^+$   & ~~$n^-$    & ~~~~~~~slope & $\chi^{2}/ndf$ \\ \hline \smallskip
 $\leq~3$  & $\leq~2$ & $-0.23\pm0.06$ & 5.7/6  \\ 
 $\leq~2$  & $\leq~1$ & $-0.25\pm0.07$ & 3.3/6 \\ 
 $\leq~2$  &       0  & $-0.30\pm0.07$ & 3.7/6 \\
       1   &       0  & $-0.37\pm0.10$ & 3.9/6 \\ \hline
\end{tabular}
\caption{\it { The fitted value of the ($\alpha, v$) slope,
and the corresponding $\chi^{2}/ndf$, for \bp
selected from events with various numbers of positive ($n^+$)
and negative ($n^-$) low  momentum ($P < 700 MeV/c$) particles.}}
\label{table:alpha}
\end{table}

\section{Conclusions}

We have observed backward proton and $\pi^-$ production in $\nu$-nucleus 
interactions in the NOMAD detector.\\
The slope parameter $B$ of the invariant cross section,
parametrized as $C~e^{-BP^{2}}$, has been measured and found to be consistent
with previous $\nu$-nucleus and hadron-nucleus experiments.
We found that $B$ does not depend on $E_{HAD}$ and $Q^{2}$ over a wide range
of values. This is in agreement with the ``nuclear scaling'' previously
observed in hadronic experiments.
The observed invariant spectrum is not consistent with the existence of 
a ``long'' Fermi momentum tail as the one proposed in Ref.~\cite{bodek} 
but agrees with the prediction of Ref.~\cite{benhar}.\\
The \bp rate in the NOMAD target (mainly carbon) has been measured 
and compared with the values obtained on different nuclei. 
While the $A$ dependence for neutrino scattering on heavy nuclei  
is consistent with that of hadron experiments, and can be parametrized
as $<N_{BP}> \propto A^{\alpha}$ with $\alpha = 0.68$ \cite{dayon}, the NOMAD result
does not fit this dependence. The $A$ dependence of \bpim
has been found to be steeper than that of $Bp$.\\
The backward proton data have been compared with the predictions of
reinteractions and short range models. The observed energy dependence
is consistent with the ``formation zone'' mechanism. The correlation
between the multiplicitiy of slow ($P<700 MeV/c$) tracks and \bp
events indicates the effects of reinteractions. However when appropriate cuts
are applied in order to reduce the intranuclear cascade contributions,
the correlation between the \bp and the muon scaling variable $v$, predicted
by the short range models, becomes stronger.

\begin{ack}
We thank the management and staff of CERN and of all
participating institutions for their vigorous support of the experiment.
Particular thanks are due to the CERN accelerator and beam-line staff
for the magnificent performance of the neutrino beam. The following
funding agencies have contributed to this experiment:
Australian Research Council (ARC) and Department of Industry, Science, and
Resources (DISR), Australia;
Institut National de Physique Nucl\'eaire et Physique des Particules (IN2P3), 
Commissariat \`a l'Energie Atomique (CEA), Minist\`ere de l'Education 
Nationale, de l'Enseignement Sup\'erieur et de la Recherche, France;
Bundesministerium f\"ur Bildung und Forschung (BMBF, contract 05 6DO52), 
Germany; 
Istituto Nazionale di Fisica Nucleare (INFN), Italy;
Russian Foundation for Basic Research, 
Institute for Nuclear Research of the Russian Academy of Sciences, Russia; 
Fonds National Suisse de la Recherche Scientifique, Switzerland;
Department of Energy, National Science Foundation (grant PHY-9526278), 
the Sloan and the Cottrell Foundations, USA. F.J.P. Soler is
supported by a TMR Fellowship from the European Commission.
We are grateful to O. Benhar, S. Fantoni and G. Lykasov for
stimulating discussions on the subject of this paper.
\end{ack}


\end{document}